\documentclass[twoside,twocolumn,english,prl,showpacs]{revtex4}
\usepackage[]{fontenc}
\usepackage[latin1]{inputenc}
\usepackage{subfigure}
\usepackage{amsmath}
\usepackage{graphicx}
\usepackage{amssymb}

\makeatletter

\providecommand{\LyX}{L\kern-.1667em\lower.25em\hbox{Y}\kern-.125emX\@}

\usepackage{graphicx}
\usepackage{txfonts}

\usepackage{babel}
\makeatother
\begin{document}

\title{Surface Phase Transitions Induced by Electron Mediated Adatom-Adatom
Interaction}

\author{Junren Shi$^{1}$, Biao Wu$^{1,\, 2}$, X.C. Xie$^{3,\, 4}$, E.W.
Plummer$^{1,\, 5}$, Zhenyu Zhang$^{1,\, 5}$}

\affiliation{$^{1}$Condensed Matter Sciences Division, Oak Ridge National Laboratory,
Oak Ridge, Tennessee 37831\\
 $^{2}$Department of Physics, University of Texas, Austin, Texas
78712\\
 $^{3}$Department of Physics, Oklahoma State University, Stillwater,
Oklahoma 74078\\
 $^{4}$International Center for Quantum Structures, Chinese Academy
of Sciences, Beijing 100080, China\\
 $^{5}$Department of Physics, University of Tennessee, Knoxville,
Tennessee 37996}

\begin{abstract}
We propose that the indirect adatom-adatom interaction mediated by
the conduction electrons of a metallic surface is responsible for
the $\sqrt{3}\times \sqrt{3}\Leftrightarrow 3\times 3$ structural
phase transitions observed in Sn/Ge~(111) and Pb/Ge~(111). When
the indirect interaction overwhelms the local stress field imposed
by the substrate registry, the system suffers a phonon instability,
resulting in a structural phase transition in the adlayer. Our theory
is capable of explaining all the salient features of the $\sqrt{3}\times \sqrt{3}\Leftrightarrow 3\times 3$
transitions observed in Sn/Ge~(111) and Pb/Ge~(111), and is in principle
applicable to a wide class of systems whose surfaces are metallic
before the transition.
\end{abstract}


\maketitle
\begin{figure}
\includegraphics[  width=0.90\columnwidth]{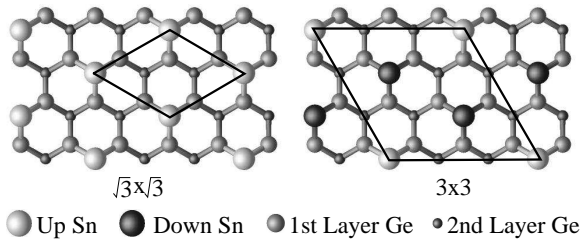}

\caption{\label{Model}Structures of $\sqrt{3}\times \sqrt{3}$ (left) and
$3\times 3$ (right) phases of Sn/Ge(111). In the experiment, one
third monoatomic layer of tin was deposited onto a Ge(111) surface,
and a gradual transition from the room-temperature flat $\sqrt{3}\times \sqrt{3}$
phase to the low-temperature $3\times 3$ phase was observed~\cite{Carpinelli1996,Carpinelli1997}.
In the new phase, one Sn adatom moves up and two move down in each
$3\times 3$ unit cell.}
\end{figure}

Over the years, a great deal of efforts have been devoted to experimental
studies of structural phase transitions at surfaces. One compelling
example is the $\sqrt{3}\times \sqrt{3}\Leftrightarrow 3\times 3$
transitions in the $1/3$ monolayer of Pb or Sn on Ge(111) directly
observed by the Scanning Tunneling Microscopy (STM)~\cite{Carpinelli1996,Carpinelli1997,Petersen2002b}
(see Fig.~\ref{Model}). These studies have stimulated an active
line of theoretical research, yet to date, the precise nature and
the underlying mechanism of such transitions are still highly controversial.
The original paper attributed this transition to a Charge Density
Wave (CDW) driven by two-dimensional Fermi surface nesting~\cite{Carpinelli1996,Carpinelli1997}.
Subsequent papers have attributed the transition to a Kohn Anomaly~\cite{Chiang2002},
bond density waves~\cite{Gironcoli2000}, a pseudo Jahn-Teller transition~\cite{Bunk1999,Zhang1999},
a surface Mott insulator~\cite{Santoro1999}, dynamical fluctuations~\cite{Avila1999},
a soft phonon~\cite{Perez2001} and most recently to disproportionation~\cite{Ballabio2002}.
Although all these theories are successful to certain extent, a unified
microscopic picture is yet to emerge. The issue becomes even more
intriguing after the observation of the delicate role of defects in
the transition~\cite{Weitering1999,Melechko2000}.

In this Letter, we present a new mechanism for surface phase transitions,
which places central emphasis on the indirect adatom-adatom interaction
mediated by the two-dimensional conduction electrons of a metallic
surface. In this theory, when the conduction-electron mediated adatom-adatom
interaction overwhelms the local stress field imposed by the substrate
registry, the system suffers a phonon instability, leading to a structural
phase transition. The theory is capable of explaining all the salient
features of the transition observed in Sn/Ge (111), including the
appearance of surface charge density waves and the delicate role of
surface defects. It also predicts the existence of a glassy phase
in such systems.

The electron-mediated adatom-adatom interaction originates from the
tendency of the conduction electrons to screen external disturbances.
In the case of adlayer systems such as Sn/Ge~(111), the dangling
bond electrons of the adatoms form a quasi two-dimensional (2D) electron
gas at the surface. The movement of an adatom disturbs the electron
gas, which responses in the form of Friedel oscillations in its density.
Such charge corrugations propagate at the surface to reach other adatoms,
thereby establishing an indirect interaction between adatoms. We note
that the basic physical concept emphasized here has found important
applications in several other fields, such as the RKKY interaction
between spins~\cite{Ruderman1954} and chemisorption of adatoms at
metal surfaces~\cite{Einstein1973,Lau1978,Repp2000}.

We stress that, the central ingredients of the picture developed here
are the existence of surface conduction electrons and their capability
of coupling with the displacement of the adatoms (electron-phonon
coupling). Both are evidently present in Sn/Ge~(111) and Pb/Ge~(111).
As shown in Fig.~\ref{Model}, each Sn adatom bonds to three Ge substrate
atoms directly under it, leaving one bond dangling. The electrons
in the dangling bonds are localized in the surface and form a 2D electron
system with a narrow ($\sim 0.5\, \mathrm{eV}$) half-filled band.
When an adatom is displaced, the angles between the dangling bond
and its three saturated Sn-Ge bonds have to adjust accordingly, inducing
a variation in the s-p hybridization of the electron states of the
Sn adatom, and a corresponding change in the energy of the dangling
state~\cite{Harrison1989}. This process can be described by the
Holstein electron-phonon coupling~\cite{Holstein1959}, where the
dangling state energy of an adatom depends on its displacement as
$\epsilon _{\mathrm{d}}=\epsilon _{0}-\beta z$, with $z$ being the
displacement of the adatom along the direction perpendicular to the
surface, and $\beta $ a positive constant, reflecting that a downward
displacement induces a higher dangling state energy. 

The total Hamiltonian of the system is written as,

\begin{equation}
H=\sum _{i,s}(\epsilon _{0}-\beta z_{i})c_{is}^{\dagger }c_{is}-t\sum _{\langle ij\rangle ,\, s}c_{is}^{\dagger }c_{js}+\frac{\alpha }{2}\sum _{i}z_{i}^{2}+\cdots ,\label{Hamiltonian}\end{equation}
 where $z_{i}$ is the displacement of the adatom located at site
$i$, and $c_{is}^{\dagger }$ ($c_{is}$) is the creation (annihilation)
operator of a dangling bond electron at site $i$ and with spin $s$.
The hopping constant $t$ is between nearest neighbors and can be
regarded as independent of the adatom displacement because the leading
correction term is second order in $z_{i}$. The term involving $\alpha $
is the elastic energy of the lattice distortion caused by the displacements
of the adatoms, which represents the local stress field imposed by
the substrate. In Eq.~\ref{Hamiltonian}, the higher order terms
in $z_{i}$ are ignored for the moment. As it will become clear in
later discussions, Eq.~\ref{Hamiltonian} is sufficient for determining
the structural stability of the system, while the higher order terms
are important only in stabilizing the system after the system becomes
unstable.

\begin{figure}
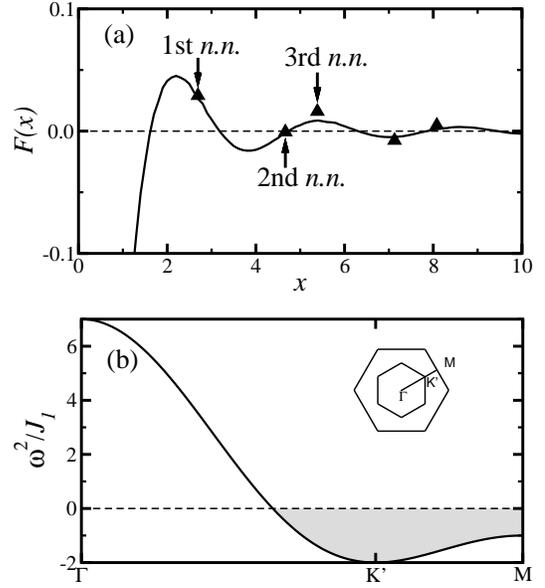

\includegraphics[  width=0.80\columnwidth]{SnGefig2a}

\vskip 3mm

\includegraphics[  width=0.80\columnwidth]{SnGefig2b}

\caption{\label{Fx_Phonon}(a) $F(x)$ in Eq.~\ref{Coeff} (as a comparison,
the triangles are the results of tight-binding calculations up to
the fifth nearest neighbors for the $\sqrt{3}\times \sqrt{3}$ structure).
In this system, the Fermi momentum $k_{F}\approx \sqrt{2\pi \rho }=\sqrt{4\pi }/(3^{1/4}a)$,
where $\rho $ is the density of the surface electrons, $a$ is the
lattice constant. (b) Phonon dispersion {[}Eq.~\ref{Classical}{]}
along the direction $\Gamma \mathrm{M}$ shown in the inset, with
$\alpha /J_{1}=7$. The shaded area indicates the unstable phonon
modes, of which the most unstable $K'$ mode defines the $3\times 3$
periodicity.}
\end{figure}

The stability of the $\sqrt{3}\times \sqrt{3}$ phase can be examined
within the perturbation theory by expanding the total energy change
up to the second order in the adatom displacements, \begin{equation}
\Delta E=\frac{\alpha }{2}\sum _{i}z_{i}^{2}-\beta \langle n\rangle \sum _{i}z_{i}-\frac{1}{4}\sum _{ij}J_{ij}(z_{i}-z_{j})^{2}+\mathcal{O}(z^{3})\, ,\label{DeltaE}\end{equation}
 where $\langle n\rangle $ is the average number of the dangling
electrons for each adatom, and $J_{ij}$ are the coupling coefficients
for the indirect adatom-adatom interaction. In the continuum limit,
we have, \begin{eqnarray}
J_{ij} & = & 8\beta ^{2}\sum _{\epsilon _{\mathbf{q}}<\epsilon _{F}}\sum _{\epsilon _{\mathbf{k}}>\epsilon _{F}}\frac{\exp \left[-i(\mathbf{q}-\mathbf{k})\cdot (\mathbf{R}_{i}-\mathbf{R}_{j})\right]}{\epsilon _{\mathbf{q}}-\epsilon _{\mathbf{k}}}\nonumber \\
 & \approx  & \frac{8n^{2}\beta ^{2}}{\epsilon _{F}}F(k_{F}|\mathbf{R}_{i}-\mathbf{R}_{j}|)\, \label{Coeff}
\end{eqnarray}
 where $\epsilon _{\mathbf{q}}$ ($\epsilon _{\mathbf{k}}$) is the
energy dispersion of the electron band, $\epsilon _{F}$ is the Fermi
energy relative to the band bottom and contains implicitly the hopping
constant $t$. The coefficients $J_{ij}$ are the same as for the
2D RKKY coupling~\cite{Fischer1975}, and are oscillatory spatially
with an asymptotic dependence of $-\sin (2k_{F}r)/r^{2}$, as shown
in Fig.~\ref{Fx_Phonon}(a). When applied to the case of $\sqrt{3}\times \sqrt{3}$
structure of Sn/Ge(111), the coupling between the nearest neighbors
is positive; therefore, the interaction is anti-ferromagnetic-like,
namely, two nearest neighboring adatoms tend to displace in opposite
directions.

Keeping only the interacting terms between nearest neighbors, we can
reduce Eq.~\ref{DeltaE} to \begin{equation}
\Delta E=\frac{1}{2}(\alpha -6J_{1})\sum _{i}\tilde{z}_{i}^{2}+\frac{J_{1}}{2}\sum _{\langle ij\rangle }\tilde{z}_{i}\tilde{z}_{j}+\mathcal{O}(\tilde{z}^{3})\, ,\label{Classical}\end{equation}
 where $\tilde{z}_{i}=z_{i}-\beta /(\alpha \langle n\rangle )$. This
equation shares the same form as the phenomenological charge compensate
model (CCM) proposed by Melechko \textit{\emph{et al.}}~\cite{Melechko2001},
which has been shown to be capable of interpreting most of the experimental
STM images. Our theory thus provides the microscopic mechanism behind
the success of the CCM model. We note that the long-range terms in
Eq.~\ref{DeltaE} should be included in future investigation of large-scale
phenomena such as domain wall formation, but they do not alter the
nature of the phase transitions, which is the focus of the present
study.

From Eq.~\ref{Classical}, it is evident that the system becomes
unstable against adatom displacements when the elastic restoring force
is weak. This can be illustrated with the phonon dispersion of Eq.~\ref{Classical}~\cite{Perez2001}:
\begin{equation}
\omega ^{2}=\alpha -6J_{1}+J_{1}\sum _{l}\cos (\mathbf{k}\cdot \mathbf{R}_{l}),\label{Dispersion}\end{equation}
 where the sum is over six nearest neighbors. When $\alpha /J_{1}<9$,
$\omega $ becomes imaginary for certain $\mathbf{k}$ vectors (Fig.~\ref{Fx_Phonon}b),
indicating that the corresponding phonon modes become unstable (phonon
catastrophe) and the system is driven into new phases. Note that the
soft phonon picture suggested in Ref.\onlinecite{Perez2001} is a
very special case of our model, occurring only when $\alpha /J_{1}$
is exactly equal to 9. Furthermore, the phonon instability established
here does not require any special properties of the Fermi surface
such as the Fermi surface nesting, nor does it rely on the inclusion
of electron-electron correlations, which is the focus of most of the
earlier efforts with little success~\cite{Santoro1999,Santoro2001}.

Higher order terms become important when the system enters the unstable
regime. On a microscopic level, those terms may arise from a variety
of sources, such as anharmonicity in the elastic energy, finite width
of the electron band, higher order terms in expansions of the site
energy and the hopping constant $t$. The precise form of these higher
order terms may affect the quantitative details of the phase transition,
such as the amplitudes of the adatom displacements and preference
among the nearly degenerate ground states, but the occurrence and
nature of the transition is totally determined by the intrinsic instability
represented in Eq.~\ref{Classical}. For this reason and also for
simplicity, we assume that the higher order terms mainly come from
the anharmonic terms of the on-site elastic energy, \begin{equation}
-\frac{\delta }{3}\sum _{i}\tilde{z}_{i}^{3}+\frac{\gamma }{4}\sum _{i}\tilde{z}_{i}^{4}\label{Nonlinearity}\end{equation}
 where $\delta >0$ and $\gamma >0$. $\delta $ is positive, because
displacing an adatom out of the surface weakens the bond strengths
between the adatom and the substrate atoms.

\begin{figure}
\subfigure[$\alpha/J_1 \geq 9$]{\includegraphics[  width=0.32\columnwidth]{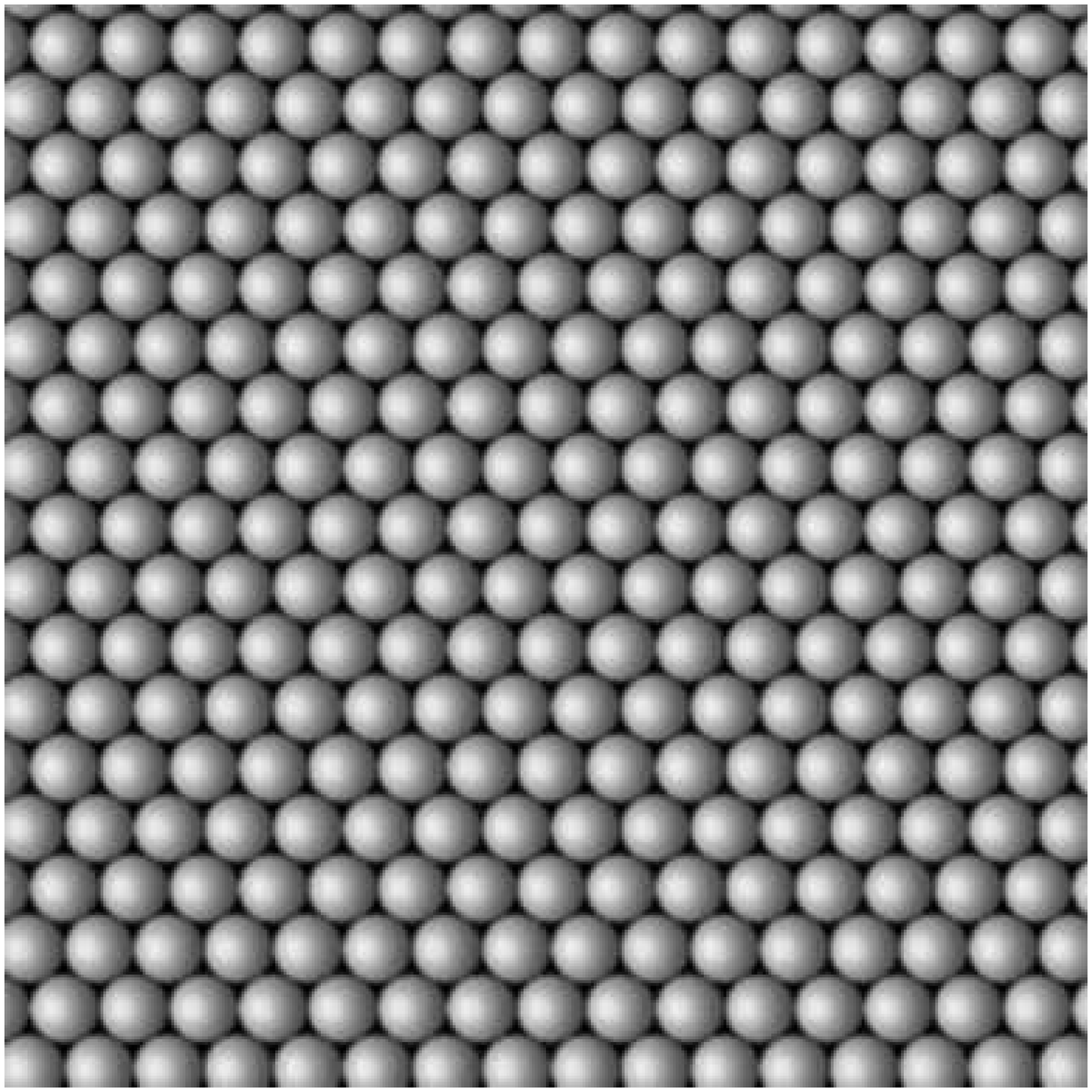}}~\subfigure[$6 \leq \alpha/J_1 <9$]{\includegraphics[  width=0.32\columnwidth]{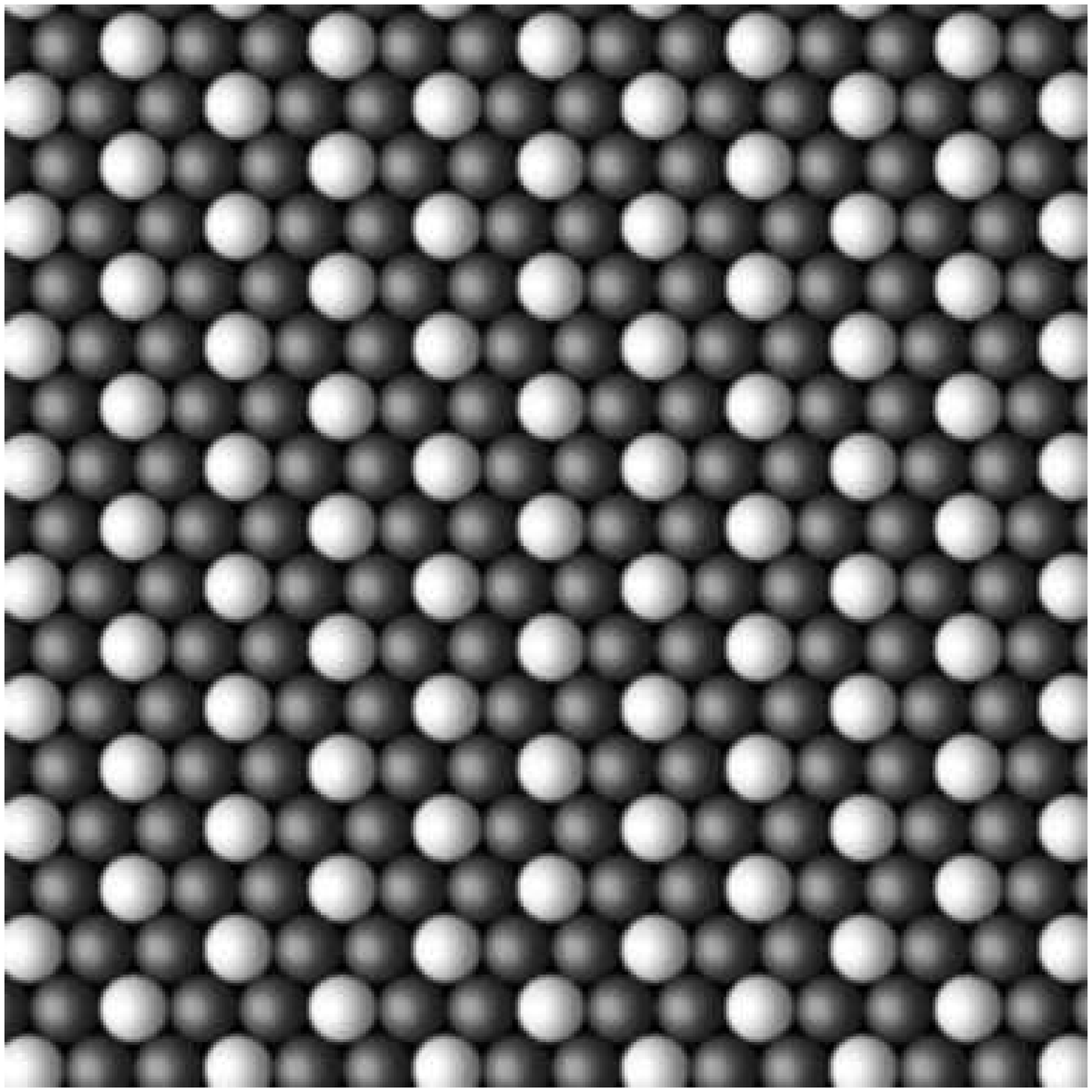}}~\subfigure[$\alpha/J_1 <6$]{\includegraphics[  width=0.32\columnwidth]{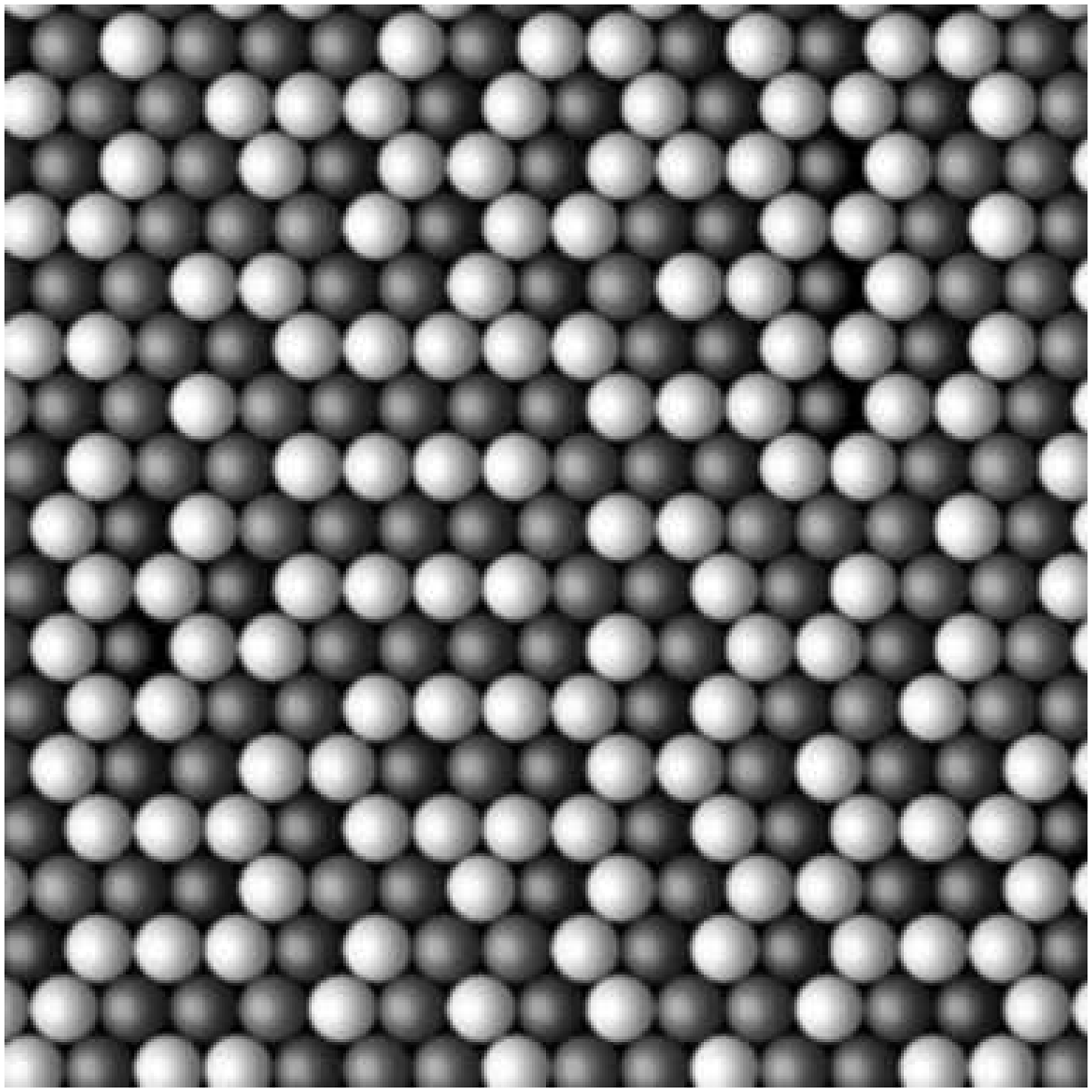}}

\caption{\label{ground_state} Schematic images of the stable configurations
for different parameter regimes. (a) $\sqrt{3}\times \sqrt{3}$ phase;
(b) $3\times 3$ phase; (c) kinked-line phase.}
\end{figure}

By adding those higher order terms into Eq.~\ref{Classical}, we
can determine the stable configurations. The numerical results are
shown in Fig.~\ref{ground_state}. The ratio $\alpha /J_{1}$ is
a crucial deterministic parameter for the phase transition. When $\alpha /J_{1}\geq 9$,
the ground state is the flat $\sqrt{3}\times \sqrt{3}$ phase. When
$6\leq \alpha /J_{1}<9$, the ground state is the $3\times 3$ phase
in the one up-two down configuration: in each unit cell of the $3\times 3$
lattice, one Sn adatom moves up ($z_{\uparrow }$), while the other
two move down ($z_{\downarrow }$), and the displacements satisfy
$z_{\uparrow }\approx 2|z_{\downarrow }|$ . The nearly degenerate
one down-two up configuration has a higher energy due to the cubic
term in Eq.~\ref{Nonlinearity}. Such $3\times 3$ patterns were
also observed experimentally~\cite{Zhang1999,Petaccia2001}. Note
that the $3\times 3$ structure corresponds to the most unstable phonon
mode in Eq.~\ref{Dispersion}, as demonstrated in Fig.~\ref{Fx_Phonon}(b).

When $\alpha /J_{1}<6$, the system shows kinked-line configuration
and behaves like a glass~\cite{Angell1995}. Starting with random
initial configurations, we always end up with a metastable disordered
structure as typified in Fig.~\ref{ground_state}(c), instead of
the true ground state $3\times 3$. We expect a similar behavior in
a realistic experiment with finite cooling rates: the system is trapped
in one of the meta-stable states, showing a disordered configuration.
Such behavior is directly related to the property of the single body
potential of an adatom ,\begin{equation}
\frac{1}{2}(\alpha -6J_{1})\tilde{z}_{i}^{2}-\frac{\delta }{3}\tilde{z}_{i}^{3}+\frac{\gamma }{4}\tilde{z}_{i}^{4}\, ,\label{eq:singlebody}\end{equation}
 which develops a double-well shape when $\alpha /J_{1}<6$. As a
result, the present model can be mapped onto an Ising antiferromagnet
by considering the displacement $\tilde{z}_{i}$ taking discrete values
at the local minima of the double-well potential. Such an Ising antiferromagnet
on a triangular lattice is known to have an exponentially large number
of degenerate ground states~\cite{Wannier1950}. Here, the vibrations
of $\tilde{z}_{i}$ around the local minima break the degeneracy,
leading to the exponentially large number of meta-stable states in
addition to its true ground state $3\times 3$. 

The dependence of the stable configurations on $\alpha /J_{1}$ qualitatively
explains why the $\sqrt{3}\times \sqrt{3}\Leftrightarrow 3\times 3$
transition is observed in the Sn/Ge and Pb/Ge systems, but not in
Sn/Si: the stronger bond strength between a Sn adatom and a substrate
Si atom places the system into the regime where $\alpha /J_{1}>9$.
This is to be verified by future first-principles calculations.

The structural phase transition manifests itself by an accompanying
charge density wave, as observed experimentally~\cite{Carpinelli1996,Carpinelli1997}.
Following the Hellmann-Feynman theorem, we obtain the total force
acting on an adatom, \begin{equation}
F_{i}=\alpha z_{i}-\beta \langle n_{i}\rangle +\mathcal{O}(z^{3})\, .\label{eq:Hellmann-Feynman}\end{equation}
 Setting $F_{i}=0$ in equilibrium, we have $\langle n_{i}\rangle =\langle n\rangle +(\alpha /\beta )\tilde{z}_{i}+\mathcal{O}(\tilde{z}^{3})$,
showing that adatoms displaced upward gain electrons while those downward
lose electrons. 

\begin{figure}
\includegraphics[  width=0.80\columnwidth]{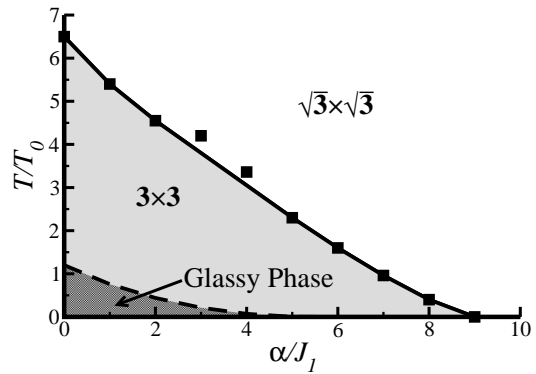}

\caption{\label{phase_diagram}Phase diagram of the system. The boundary between
the glassy phase and the $3\times 3$ ground state is not well defined
as indicated with the dashed line. $T_{0}=J_{1}^{2}/\gamma $, $\delta =0$.}
\end{figure}

\begin{figure}
\includegraphics[  width=0.80\columnwidth]{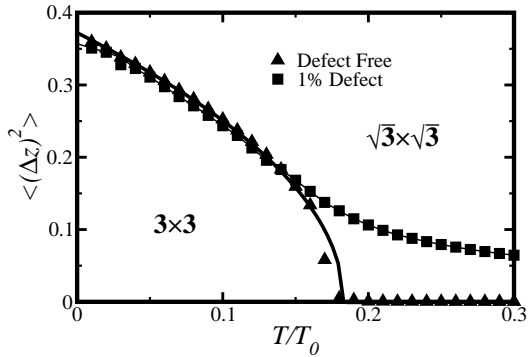}

\caption{\label{Order_Parameter}Temperature dependence of the order parameter
for the defect-free system and the system with $1\%$ defect. The
thick solid line shows the function $A(1-T/T_{c})^{1/2}$ with $T_{c}/T_{0}=0.18$,
$T_{0}=J_{1}^{2}/\gamma $. The parameters are $\alpha /J_{1}=8.5$,
$J_{1}=1$, $\gamma =1.0$, and $\delta =0.1$, with a simulation
system size of $99\times 99$.}
\end{figure}

We now study the finite temperature behavior of the system by applying
the Monte-Carlo algorithm to the effective classical Hamiltonian shown
in Eq.~\ref{Classical} along with Eq.~\ref{Nonlinearity}. We find
that the transition between the $\sqrt{3}\times \sqrt{3}$ phase and
the $3\times 3$ phase is of second-order with a sharply defined boundary,
as shown in Fig.~\ref{phase_diagram}. There exists another boundary
between the glassy phase and the $3\times 3$ phase, which is the
result of the glassy states at $T=0$ when $\alpha /J_{1}<6$. As
is typical in a glassy system, the boundary of the glassy phase is
not well defined. The detailed behavior of the $\sqrt{3}\times \sqrt{3}\Leftrightarrow 3\times 3$
phase transition along the temperature axis is shown in Fig.~\ref{Order_Parameter}.
The order parameter is chosen as the mean square corrugation of the
thermo-average positions of adatoms~\cite{Melechko2001}. The defect-free
system clearly shows a second-order transition from the $3\times 3$
phase to the $\sqrt{3}\times \sqrt{3}$ phase as seen in Fig.~\ref{Order_Parameter},
where the temperature dependence of the order parameter is well fitted
by $A(1-T/T_{c})^{1/2}$. This behavior is different from the prediction
of the Ginzburg-Landau type theories such as the CCM, which concludes
that the critical behavior follows $|1-T/T_{c}|$~\cite{Melechko2001}.

Figure~\ref{Order_Parameter} also shows that the sharp phase transition
in a pure system is blurred by the presence of a very low concentration
of substitutional defects. In our calculations, the defect is simulated
by introducing a constant displacement, \begin{equation}
\Delta z=-\Delta \epsilon /\beta \, .\label{eq:ZD}\end{equation}
 where $\Delta \epsilon $ is the site energy difference between a
Ge substitutional defect and a Sn adatom~\cite{Chiang2002}. As seen
in Fig.~\ref{Order_Parameter}, with only $1\%$ defects, the sharp
transition in the order parameter converts into a crossover behavior,
explaining the experimental observation that the transition is gradual.
Just like in the experiments~\cite{Weitering1999,Melechko2001},
our simulations show that each defect induces a local $3\times 3$
patch above $T_{c}$, and the size of each individual $3\times 3$
patch decreases gradually with the temperature.

In conclusion, we have presented a complete theory to understand the
surface phase transitions observed in Sn/Ge~(111) and Pb/Ge~(111).
Although we focus on these specific systems, the central ingredients
of the theory, namely, the electron mediated indirect interaction
and the resulting phonon instability, are conceptually much more general.
These basic ideas, once properly applied, will enable us to gain better
understanding of structural phase transitions in a wide class of systems
whose surfaces are metallic before the transition.

We gratefully acknowledge valuable discussions with A. Melechko, H.
Weitering, Ismail, A. Chernyshev and Q. Niu. This work was supported
in part by the LDRD of ORNL, managed by UT-Battelle, LLC for the USDOE
(DE-AC05-00OR22725). It was also supported by the the NSF DMR-0105232
(EWP), DMR-0071893 (BW and ZZ) and the USDOE DE-FG03-01ER45687 (XCX).

\bibliographystyle{apsrev}
\bibliography{Refs}

\end{document}